\documentclass{svproc}
\usepackage[table,dvipsnames]{xcolor}
\usepackage{amsmath,amsfonts,graphicx,hyperref,tikz-network,tikz,booktabs,cleveref,enumitem,csquotes,bold-extra,multirow,array,pgfplots} 

\pgfplotsset{compat=1.11}

\newcolumntype{C}[1]{>{\centering\let\newline\\\arraybackslash\hspace{0pt}}m{#1}}
\renewcommand{\(}{\left(}
\renewcommand{\)}{\right)}
\newcommand{\qt}{\enquote}
\renewcommand{\b}{\bfseries}
\renewcommand{\phi}{\varphi}
\newcommand{\D}{\mathrm\Delta}
\newcommand{\voc}{v_{\text{oc}}}
\begin{document}
\mainmatter
\title{Oceanic Games: Centralization Risks and Incentives in Blockchain Mining}
\titlerunning{Oceanic Games in Blockchain Mining}
\author{Nikos Leonardos\inst{1} \and
Stefanos Leonardos\inst{2} \and
Georgios Piliouras\inst{2}}
\authorrunning{N. Leonardos et al.}
\institute{National and Kapodistrian University of Athens, Panepistimioupolis, Zografou 161 22, Greece.
\email{nleon@di.uoa.gr} \and
Singapore University of Technology and Design, 8 Somapah Rd, 487372, Singapore.
\email{\{stefanos\_leonardos,georgios\}@sutd.edu.sg}}

\maketitle
\begin{abstract}
To participate in the distributed consensus of permissionless blockchains, prospective nodes -- or \emph{miners} -- provide proof of designated, costly resources. However, in contrast to the intended decentralization, current data on blockchain mining unveils increased concentration of these resources in a few major entities, typically \emph{mining pools}. To study strategic considerations in this setting, we employ the concept of \emph{Oceanic Games} \cite{Mi78}. Oceanic Games have been used to analyze decision making in corporate settings with small numbers of dominant players (shareholders) and large numbers of individually insignificant players, \emph{the ocean}. Unlike standard equilibrium models, they focus on \emph{measuring the value (or power) per entity and per unit of resource} in a given distribution of resources. These values are viewed as strategic components in coalition formations, mergers and resource acquisitions. Considering such issues relevant to blockchain governance and long-term sustainability, we adapt oceanic games to blockchain mining and illustrate the defined concepts via examples. The application of existing results reveals incentives for individual miners to merge in order to increase the value of their resources. This offers an alternative perspective to the observed centralization and concentration of mining power. Beyond numerical simulations, we use the model to identify issues relevant to the design of future cryptocurrencies and formulate prospective research questions.     
\end{abstract}

\section{Introduction}
\label{sec:introduction}
Decentralization is a core element in the design of permissionless blockchains \cite{Pre20}. To participate in the blockchain consensus mechanisms, prospective network nodes -- also called \emph{miners} -- need to provide proof of some costly resource. This resource may be computational power in protocols with Proof of Work (PoW) selection mechanisms, \cite{Na08,Ga15}, or coins of the native cryptocurrency in Proof of Stake (PoS) selection mechanisms, \cite{Bo15,Be16}. Under default conditions, the selection is proportional to miners' resources and hence, it depends on their actual distribution. An integral assumption in the security philosophy of permissionless blockchains is that the network of mining nodes remains \qt{sufficiently} decentralized and distributed. In the extreme case, \emph{sufficiently} means that no single entity holds $50\%$ or more of the resources but in practice much more fragmentation may be desired to safeguard the safety properties of the underlying protocol \cite{Ey14,Ki16,Bad18}.\par
With this in mind, the picture illustrated in \Cref{tab:hashpower} is disconcerting.
\begin{table}[!htb]
\vspace{-0.4cm}\centering
\begin{tabular}{llrllr}
\toprule
& \multicolumn{2}{c}{\bfseries{\scshape{Bitcoin}}} &\phantom{3333}\hspace{50pt}& \multicolumn{2}{c}{\bfseries{\scshape{Ethereum}}}\\ \cmidrule{1-3}\cmidrule{5-6}
& \b Entity (Pool) &\b Blocks (\%) && \b Entity (Pool) & \b Blocks (\%)\\
\cmidrule{1-3}\cmidrule{5-6}
1. & BTC.com & $18.2$ && Ethermine & $28.2$\\
2. & AntPool & $14.7$ && Sparkpool & $21.4$\\
3. & F2Pool & $12.6$ && Nanopool & $12.6$\\
4. & SlushPool & $10.1$ && F2Pool\_2 & $12.4$\\
5. & BTC.TOP & $7.9$ && MiningPoolHub\_1 & $5.6$\\
6. & ViaBTC & $7.9$ && DwarfPool\_1 & $1.9$\\
7. & DPOOL & $4.1$ && PandaMiner & $1.8$\\
8. & BitFury & $2.3$ && firepool & $1.6$\\
9. & BitClub Network & $2.3$ && Address\_1 & $1.4$\\
10. & Bitcoin.com & $1$ && MinerallPool & $1.1$\\
& \b Unknown/other & \b 18.9 && \b Unknown/other& \b 12.0 \\
\bottomrule
\end{tabular}\\[0.4cm]
\caption{Distribution of the blocks mined in the Bitcoin and Ethereum blockchains. Mining is dominated by few major miners, typically mining pools, numbered from 1 to 10 and a great number of minor players in the \qt{Unkown/other} category. Source: \href{https://www.blockchain.com/en/pools?timespan=4days}{blockchain.com} and \href{https://etherscan.io/stat/miner?blocktype=blocks}{etherscan.io}, 5 March 2019.}
\label{tab:hashpower}\vspace{-0.7cm}
\end{table}
\Cref{tab:hashpower} shows the distribution of blocks among miners in the two largest\footnote{In terms of market capitalization, cf. \href{https://coinmarketcap.com/}{coinmarketcap.com}.} cryptocurrencies, Bitcoin and Ethereum, and indicates that the desired assumption of a highly decentralized (and distributed) network is currently not satisfied. As can be seen, the vast majority of mining resources is concentrated in a small number of \qt{major} nodes or \emph{mining pools} in which individual miners join forces to reduce the variance of their payments \cite{Fi17,So18}. The rest is scattered among a large number of minor and individually insignificant miners. The discrepancy between the intended distribution and the concentration of resources that is observed in practice raises some questions. What is the actual power of such pools or major miners to influence the evolution of the blockchain? Does this distribution create incentives for mergers and formation of coalitions (cartels) that will seize control of the majority of resources and manipulate the blockchain \cite{Le15,La15}? What strategic considerations arise and what are their implications on blockchain governance and long-term sustainability? \par
Similar questions have been examined by conventional economics in the context of corporate governance. To study interactions between shareholders with various degrees of power in particular, \cite{Mi78} developed the model of \emph{Oceanic Games}. These are games featuring a mixture of few large players (shareholders) and a continuum of infinitesimal players, called the \emph{ocean}, each of which holds an insignificant fraction of corporate shares. The resemblance with blockchain mining -- with shares corresponding to units of mining resources -- is apparent. Our goal in this paper is to explore incentives in blockchain mining from the perspective of Oceanic Games and complement existing studies that focus on safety and security related issues \cite{Nay16,Fa18,Brj18}.\par
The central idea in the literature of Oceanic Games is the measurement of a \emph{value} for each entity and for each \emph{unit of resource} given the distribution of resources among shareholders. The concept of \emph{value} is considered as a powerful tool in the theory of decision making \cite{Au64,Sh73,Sh78} and \cite{Sh88}. For instance, if a miner holds $51\%$ of the total resources, then each of her units is worth much more than if she holds only $49\%$ of the total resources, since in the former case, the entirety of her shares gives her absolute control over the blockchain. Similar, but maybe less obvious considerations, arise also in intermediate cases. If a miner holds $49\%$ of the resources and a second miner holds $2\%$ of the resources, then both miner's resources value higher than in the case in which the first miner only holds $47\%$ of the resources, since in the former case, the two miners may collude and jointly seize control of the blockchain. \par
Motivated by these considerations, we adapt the model of Oceanic Games from \cite{Mi78} on blockchain mining. Our aim is to measure the \emph{value of mining resources per miner and per unit of resource} as a strategic component in the process of power gain and coalition formation between mining nodes. With this approach, we shift our attention from safety attacks and equilibration models, \cite{Jo14,Ey15}, to the understanding of incentives related to the distribution and acquisition of protocol resources. The analysis of these issues is relevant to the broader subjects of long term sustainability and blockchain governance \cite{Bo16}. 
\par
Based on the above, our contribution in the present paper can be summarized in the following points 
\begin{itemize}[leftmargin=*]
\item We model instances of blockchain mining as Oceanic Games: the discrete set of large players corresponds to the large mining nodes, typically mining pools,  and the continuum of infinitesimal oceanic players to the remaining, individual miners, cf. \Cref{fig:decentralization}. 
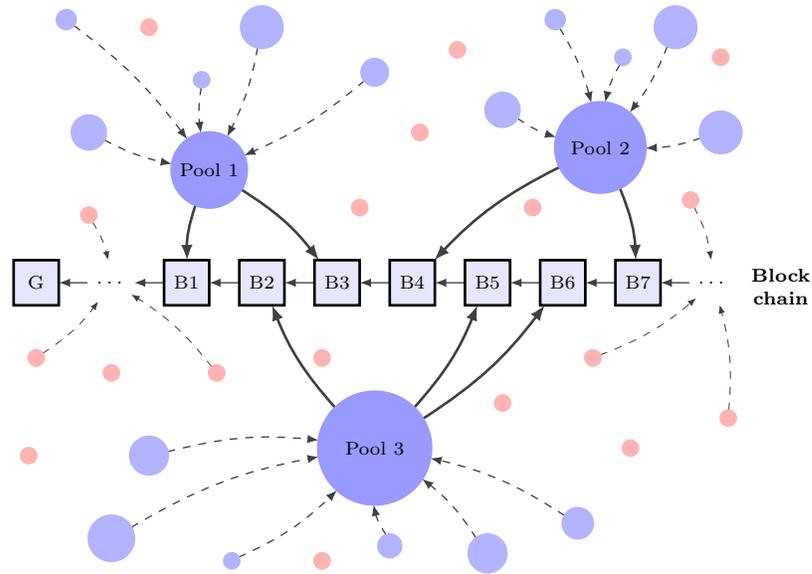
\begin{figure}[!htb]
\vspace{0cm}\centering
\begin{tikzpicture}
\clip (-0.4,0) rectangle (11.5,7.8);
\Plane[x=1,y=-1, width=1,height=11,color=white, style={dashed,inner color=white}, NoBorder, opacity=0]
\Vertex[x=1,y=4,shape=rectangle, label=G, color=blue!10]{g}
\Vertex[x=2,y=4,style={color=white},opacity=0,label=$\dots$]{dots1}
\Vertex[x=3,y=4,shape=rectangle, label=B1, color=blue!10]{b1}
\Vertex[x=4,y=4,shape=rectangle, label=B2, color=blue!10]{b2}
\Vertex[x=5,y=4,shape=rectangle, label=B3, color=blue!10]{b3}
\Vertex[x=6,y=4,shape=rectangle, label=B4, color=blue!10]{b4}
\Vertex[x=7,y=4,shape=rectangle, label=B5, color=blue!10]{b5}
\Vertex[x=8,y=4,shape=rectangle, label=B6, color=blue!10]{b6}
\Vertex[x=9,y=4,shape=rectangle, label=B7, color=blue!10]{b7}
\Vertex[x=10,y=4,style={color=white},opacity=0,label=$\dots$]{dots2}
\Edge[,lw=0.5,Direct](dots1)(g)
\Edge[,lw=0.5,Direct](b1)(dots1)
\Edge[,lw=0.5,Direct](b2)(b1)
\Edge[,lw=0.5,Direct](b3)(b2)
\Edge[,lw=0.5,Direct](b4)(b3)
\Edge[,lw=0.5,Direct](b5)(b4)
\Edge[,lw=0.5,Direct](b6)(b5)
\Edge[,lw=0.5,Direct](b7)(b6)
\Edge[,lw=0.5,Direct](dots2)(b7)
\Text[x=10.9,y=4.1,fontsize=\scriptsize]{\b Block}
\Text[x=10.9,y=3.8,fontsize=\scriptsize]{\b chain}

\Vertex[x=1.4,y=7.5,size=0.25,style={color=blue!30}]{m1}
\Vertex[x=1.7,y=6,size=0.45,style={color=blue!30}]{m2}
\Vertex[x=3.2,y=6.7,size=0.2,style={color=blue!30}]{m3}
\Vertex[x=4,y=7.4,size=0.55,style={color=blue!30}]{m4}
\Vertex[x=5.5,y=6.8,size=0.35,style={color=blue!30}]{m5}
\Vertex[x=3.3,y=5.5,size=1,style={color=blue!40}, label=Pool 1]{pool1}
\Edge[,lw=0.5,bend=8.531,Direct, style={dashed}](m1)(pool1)
\Edge[,lw=0.5,bend=-8.531,Direct, style={dashed}](m2)(pool1)
\Edge[,lw=0.5,bend=-8.531,Direct, style={dashed}](m3)(pool1)
\Edge[,lw=0.5,bend=8.531,Direct, style={dashed}](m4)(pool1)
\Edge[,lw=0.5,bend=8.531,Direct, style={dashed}](m5)(pool1)
\Edge[,lw=1,bend=10,Direct](pool1)(b3)
\Edge[,lw=1,bend=-10,Direct](pool1)(b1)

\Vertex[x=7.9,y=7.5,size=0.25,style={color=blue!30}]{m6}
\Vertex[x=7.2,y=6.3,size=0.45,style={color=blue!30}]{m7}
\Vertex[x=8.8,y=7,size=0.2,style={color=blue!30}]{m8}
\Vertex[x=9.5,y=7.4,size=0.55,style={color=blue!30}]{m9}
\Vertex[x=10.1,y=6,size=0.55,style={color=blue!30}]{m10}
\Vertex[x=8.5,y=5.8,size=1.2,style={color=blue!40}, label=Pool 2]{pool2}
\Edge[,lw=0.5,bend=8.531,Direct,style={dashed}](m6)(pool2)
\Edge[,lw=0.5,bend=-8.531,Direct,style={dashed}](m7)(pool2)
\Edge[,lw=0.5,bend=-8.531,Direct,style={dashed}](m8)(pool2)
\Edge[,lw=0.5,bend=8.531,Direct,style={dashed}](m9)(pool2)
\Edge[,lw=0.5,bend=8.531,Direct,style={dashed}](m10)(pool2)
\Edge[,lw=1,bend=-10,Direct](pool2)(b4)
\Edge[,lw=1,bend=10,Direct](pool2)(b7)

\Vertex[x=2.5,y=1.7,size=0.5,style={color=blue!30}]{u1}
\Vertex[x=3.6,y=0.3,size=0.2,style={color=blue!30}]{u2}
\Vertex[x=2,y=0.6,size=0.6,style={color=blue!30}]{u3}
\Vertex[x=5.7,y=0.5,size=0.3,style={color=blue!30}]{u4}
\Vertex[x=7,y=0.4,size=0.5,style={color=blue!30}]{u5}
\Vertex[x=8.2,y=0.8,size=0.4,style={color=blue!30}]{u6}
\Vertex[x=5.5,y=1.8,size=1.5,style={color=blue!40}, label={Pool 3}]{pool3}
\Edge[,lw=0.5,bend=10,Direct, style={dashed}](u1)(pool3)
\Edge[,lw=0.5,bend=-10,Direct, style={dashed}](u2)(pool3)
\Edge[,lw=0.5,bend=10,Direct, style={dashed}](u3)(pool3)
\Edge[,lw=0.5,bend=10,Direct, style={dashed}](u4)(pool3)
\Edge[,lw=0.5,bend=-10,Direct, style={dashed}](u5)(pool3)
\Edge[,lw=0.5,bend=-10,Direct, style={dashed}](u6)(pool3)
\Edge[,lw=1,bend=10,Direct](pool3)(b2)
\Edge[,lw=1,bend=-10,Direct](pool3)(b5)
\Edge[,lw=1,bend=-10,Direct](pool3)(b6)

\Vertex[x=0.9,y=1.7,size=0.1,style={color=red!30}]{o2}
\Vertex[x=1,y=3,size=0.2,style={color=red!30}]{o3}
\Vertex[x=1.7,y=4.9,size=0.1,style={color=red!30}]{o4}
\Vertex[x=2,y=2.8,size=0.2,style={color=red!30}]{o5}
\Vertex[x=2.5,y=7.4,size=0.1,style={color=red!30}]{o6}
\Vertex[x=3.4,y=2.8,size=0.2,style={color=red!30}]{o7}
\Vertex[x=4.8,y=3,size=0.1,style={color=red!30}]{o8}
\Vertex[x=5.3,y=5,size=0.1,style={color=red!30}]{o10}
\Vertex[x=6.1,y=6,size=0.2,style={color=red!30}]{o11}
\Vertex[x=6.6,y=7.1,size=0.1,style={color=red!30}]{o12}
\Vertex[x=8.9,y=1.8,size=0.1,style={color=red!30}]{o13}
\Vertex[x=7.2,y=2.4,size=0.1,style={color=red!30}]{o14}
\Vertex[x=7.6,y=5,size=0.1,style={color=red!30}]{o15}
\Vertex[x=8.4,y=3,size=0.2,style={color=red!30}]{o17}
\Vertex[x=4.8,y=0.3,size=0.2,style={color=red!30}]{o18}
\Vertex[x=9.7,y=5.1,size=0.1,style={color=red!30}]{o20}
\Vertex[x=10.1,y=7,size=0.1,style={color=red!30}]{o21}
\Vertex[x=10.2,y=2.2,size=0.1,style={color=red!30}]{o22}
\Edge[,lw=0.2,bend=-10,Direct,style={dashed}](o3)(dots1)
\Edge[,lw=0.2,bend=-10,Direct,style={dashed}](o22)(dots2)
\Edge[,lw=0.2,bend=10,Direct,style={dashed}](o4)(dots1)
\Edge[,lw=0.2,bend=-10,Direct,style={dashed}](o17)(dots2)
\Edge[,lw=0.2,bend=10,Direct,style={dashed}](o20)(dots2)
\Edge[,lw=0.2,bend=-10,Direct,style={dashed}](o7)(dots1)
\end{tikzpicture}
\caption{Illustration of centralization in blockchain mining. Miners join forces in few major mining pools (blue), $M=\{1,2,\dots,m\}$, which dominate the mining process. The remaining small miners (light red) -- or the ocean, $I$ -- mine individually.}\vspace{-0.5cm}
\label{fig:decentralization}
\end{figure}
Conveniently, the resulting model does not depend on the underlying selection mechanism (PoW, PoS or similar) or consensus protocol and hence can be used for the study of resource acquisition, strategic interactions, coalition formations (mergers) and governance related issues in a broad spectrum of permissionless blockchains \cite{Ga17,Brj18,thunder18,ouro17,algo17,Ki18,Bu19}. We extend an example of \cite{Mi78} to illustrate the defined concepts in blockchain context.
\item The application of existing results uncovers incentives for the formation of mergers between miners. Starting from an initial distribution in which the oceanic players control the majority of resources, we use simulations to show that this holds in two instances: first, in the formation -- crystallization -- of a coalition out of the ocean and second, in the exogenous acquisition of additional resources by a group of individual miners who, nevertheless, have the ability to coordinate their actions (collude). In both cases, the value of the miners' resources is higher when they act as a single entity rather than individual, oceanic players. This result provides an alternative perspective to the observed centralization in cryptocurrency mining, cf. \Cref{tab:hashpower}. 
\item Further numerical simulations demonstrate that the above conclusions do not hold in the whole range of parameters. Instead, the dynamics of coalition formations and entry barriers are shown to depend on the current distribution of mining resources among major miners and the ocean. 
\item Finally, we use this model to raise issues relevant to the design of future cryptocurrencies and formulate prospective research questions.
\end{itemize}
In general, the present paper can be seen as a first step towards the application of the Oceanic Game concept in blockchain mining. Beyond some first insight, the extent to which this model can provide further results in the issues of (de)centralization, blockchain governance and long-term sustainability is yet to be fully understood. 

\subsection{Outline}
\label{sub:related_work}
The rest of the paper is structured as follows. In \Cref{sec:model} we present the model of Oceanic Games and give an example to illustrate the defined notions. Revelant results from \cite{Mi78} and their application in blockchain settings are shown in \Cref{sec:results} along with numerical simulations. In \Cref{sub:research}, we raise related issues and research questions and discuss limitations of the current approach. \Cref{sec:conclusions} concludes the paper.

\section{The Model: Oceanic Games on the Blockchain}
\label{sec:model}
The current model adjusts the notation and terminology of \cite{Mi78} in a standard blockchain setting.
\begin{description}[leftmargin=0cm,itemsep=0.15cm]
\item[Miners:] The \emph{miners} are the physical entities that participate in the block proposal and creation process. The term is used here in the broadest sense and depending on the underlying protocol and selection mechanism, it may refer to \qt{conventional} miners as in PoW, \cite{Na08}, or to \emph{virtual miners} as in PoS or other alternative forms \cite{Bo15}. The set of miners consists of two distinctive components

\begin{itemize}[topsep=5pt, leftmargin=*]
\item A finite, discrete set $M=\{1,2,\dots,m\}$ of major miners or mining pools. 
\item An interval $I=[0,1]$ of infinitesimal miners. We refer to $I$ as the \emph{ocean} and to miners in $I$, as \emph{oceanic players}. We only consider subsets $U=[u_1,u_2]\subseteq I$ of the ocean $I$, e.g. $U=[0.1,0.5]$, and not individual oceanic players.  
\end{itemize}
\item[Resources:] To participate in the distributed consensus, each miner needs to provide proof of some designated, costly resource. This may be a physical or digital asset such as computational power in PoW or native coins in PoS mechanisms, respectively. To describe these resources, we use following notation

\begin{itemize}[topsep=5pt,leftmargin=*]
\item A set of real numbers $r_1,r_2,\dots,r_m \ge0$, where $r_i$ denotes the amount of resources of miner $i\in M$. For any subset $S\subseteq M$, we will write $r\(S\)=\sum_{i\in S}r_i$ to denote the total resources of miners in $S$.
\item A positive constant $\alpha>0$ which denotes the total resources of the ocean $I$. Accordingly, any subset $U=[u_1,u_2]\subseteq I$ controls $\alpha \cdot |U|$ of resources where $|U|=u_2-u_1$.
\end{itemize}
Based on the above, the total protocol resources $R$ are equal to $R:=\alpha+\sum_{i\in M}r_i$. While resources change over time, in the present analysis, we will focus on a single period or a static setting and hence, unless indicated otherwise, our notation is independent of the time $t$. Resources may be expressed as absolute numbers or percentages but this will be made explicitly clear from the context. 
\item[Blockhain Oceanic Games:] Given the above, a \emph{blockchain oceanic game} $\Gamma$ is defined by a \emph{majority quota}, $q\ge 0$, using the symbol\footnote{The notation is common in the literature of \emph{weighted voting games}, see \cite{Sh78,Sh88} and \cite{Le19} for a more related application. Also, in most cases, we will be interested in $q=0.5$ or $50\%$ but the current model applies to any $q$ of interest.} 
\[\Gamma:=\left[q; r_1,r_2,\dots, r_m;\alpha\right]\]
with the following interpretation: a coalition of miners $C:=S\cup U$ with $S\subseteq M$ and $U\subseteq I$ wins in the game $\Gamma$, if and only if its total resources are larger than or equal to $q$, i.e., if \[r\(C\):=r\(S\)+\alpha\cdot |U|\ge q\]
\item[Addition of resources:] Given an oceanic game $\Gamma$, we want to study the situation in which new entities acquire resources and enter the protocol. For this, we will use the notation $\Gamma^+$ with 
\[\Gamma^+:=\left[q; r_1,r_2,\dots, r_m,r_{m+1};\alpha\right]\]
and $M^+=\{1,2,\dots,m,m+1\}$. In words, $\Gamma^+$ results from $\Gamma$ by the addition of a new major player $m+1$ with exogenous resources $r_{m+1}>0$. Similar notation can be used to denote the formation of a new entity \emph{crystallizing} out of the ocean. In this case, $\Gamma^+:=\left[q; r_1,r_2,\dots, r_m,r_{m+1};\alpha-r_{m+1}\right]$ for some $r_{m+1}>0$, and $M=\{1,2,3,\dots, m+1\}$. 
\item[Values:] The first core functionality of the present model is to calculate a \emph{value} $\phi_i$ for each major miner $i\in M$ and one value $\Phi$ for the entirety of the oceanic players, also referred to as the \emph{oceanic value}. Each miner's value depends on that miner's share of resources \emph{and} on the total distribution of the remaining resources among the rest of major and oceanic miners. To define the miner's values $\phi_i, i \in M$ and the oceanic value $\Phi$, let $X_1,X_2,\dots,X_m$ be independent random variables uniformly distributed on $I=[0,1]$. For each $x\in I$, let $r\(x\):=\sum_{j\in M}r_j\cdot \mathbf 1\{X_j<x\}$, where $\mathbf 1\{X_j<x\}=1$ if $X_j<x$ and $0$ otherwise (indicator function). Then, the \emph{value} of miner $i\in M$ is defined by
\begin{equation}\label{eq:values}\phi_i:=\mathbb P\text{rob}\left[r\(X_i\)+\alpha X_i<q\le r\(X_i\)+\alpha X_i+r_i\right]\end{equation}
and the oceanic value by $\Phi:=1-\sum_{i\in M}\phi_i$. Intuitively, the value $\phi_i$ is the probability that miner $i$ will be the crucial entity to turn a random coalition of miners from losing (total resources of the coalition without $i$ are less than $q$) to winning (total resources of the coalition with $i$ are equal to or greater than $q$)\footnote{For more details and the probabilistic derivation of these values, we refer to \cite{Mi78}.}. 
\item[Value-per-unit of resource:] The second functionality of this model is to determine the \emph{value per-unit of resource} or power ratio, $v_i$, for each player $i\in M$, which is defined by
\begin{equation}\label{eq:ratio} v_i:=\phi_i/r_i \end{equation}
Similarly, the \emph{value per oceanic unit of resource} or oceanic power ratio, $\voc$, is equal to $\voc:=\Phi/\alpha$.
\end{description}

\subsection{An Example: Why Values and not Shares?}
\label{sub:example}
We illustrate the above with the help of an example adapted from \cite[Section 6]{Mi78}. We consider a mining situation with two major mining entities or pools, $M=\{1,2\}$, and the simple majority quota $q=0.5$ represented by the following game $\Gamma=\left[0.5; r_1, r_2; \alpha\right]$, where $\alpha=1-r_1-r_2$. The 0.5 or $50\%$ quota corresponds to cntrol of the majority of protocol resources and hence, of the blockchain as a whole. In this game a coalition $S$ wins, if $r\(S\)\ge 0.5$, i.e., if it occupies $50\%$ or more of the protocol resources\footnote{Due to continuity properties, there is no difference between using the $q=50\%$ quota or symbolically, the $q=51\%$ quota, as is common in the related literature \cite{Na08,Ey14,Ki16}.}. \par
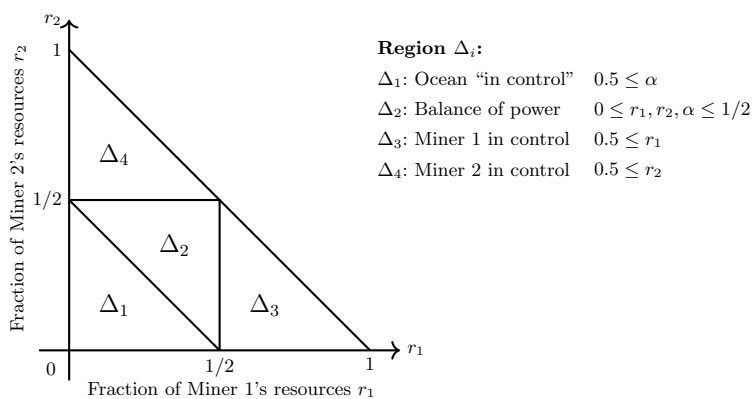
\begin{figure}[!htp] \centering
\vspace{-0.4cm}\begin{tikzpicture}[scale=0.8,thick,every node/.style={transform shape}]
\draw[->] (-0.5,0) -- (5.5,0) node[right] {$r_1$}; 
\draw[->] (0,-0.5) -- (0,5.5) node[left] {$r_2$};
\draw (0,5) node[left] {1} -- (5,0) node[below] {1};
\draw (2.5,0) node[below] {$1/2$} -- (2.5,2.5) -- (0,2.5) node[left] {$1/2$};
\draw (0,2.5) -- (2.5,0);
\node at (-0.3,-0.3) {$0$};
\node[below] at (2.7,-0.4) {Fraction of Miner $1$'s resources $r_1$};
\node[below,rotate=90] at (-1.1,2.7) {Fraction of Miner $2$'s resources $r_2$};
\node (delta1) at (0.75,0.75) {\large$\D_1$};
\node (delta2) at (1.75,1.75) {\large$\D_2$};
\node (delta3) at (3.25,0.75) {\large$\D_3$};
\node (delta4) at (0.75,3.25) {\large$\D_4$};

\node[anchor=west] (legend) at (5,5) {\b Region $\D_i$:};
\node[anchor=west] (entry1) at (5,4.5) {$\D_1$: Ocean \qt{in control}};
\node[anchor=west] (entry2) at (5,4) {$\D_2$: Balance of power};
\node[anchor=west] (entry3) at (5,3.5) {$\D_3$: Miner 1 in control};
\node[anchor=west] (entry4) at (5,3) {$\D_4$: Miner 2 in control};
\node[anchor=west] (entry11) at (8.6,4.5) {$0.5\le\alpha$};
\node[anchor=west] (entry21) at (8.6,4) {$0\le r_1,r_2,\alpha \le 1/2$};
\node[anchor=west] (entry31) at (8.6,3.5){$0.5\le r_1$};
\node[anchor=west] (entry41) at (8.6,3) {$0.5\le r_2$};
\end{tikzpicture}
\caption{All possible configurations in the distribution of resources $\(r_1,r_2\)$ between 2 major miners and the ocean, $\alpha$.}
\label{fig:example} \vspace{-0.4cm}
\end{figure}

All possible resource configurations $\(r_1,r_2,\alpha\)$ are illustrated in \Cref{fig:example}. The horizontal and vertical axes represent miner $1$'s and miner $2$'s fraction of the resources, respectively. Their possible combinations are divided in $4$ inner regions, $\D_i, i=1,2,3,4$. Region $\D_1$ contains all configurations for which the combined resources of both major miner are less than $50\%$, i.e., $r_1+r_2\le 0.5$. In this case, the majority of resources is controlled by oceanic players. However, the ocean is not actually \qt{in control}, since, by assumption, there is no coherence nor organizational structure between oceanic players. The explanation of regions $\D_2,\D_3$ and $\D_4$ is similar and is briefly given in the legend of \Cref{fig:example}.

\noindent Using \eqref{eq:values}, the value $\phi_1$ of the first major miner is given by\vspace{-0.2cm}
\begin{equation}\label{ex:value}
\phi_1=\begin{cases}
\frac{r_1\bar{r}_2}{\alpha^2},& \text{if } \(r_1,r_2\)\in \D_1\\
\(\frac{1-2r_2}{2\alpha}\)^2,& \text{if } \(r_1,r_2\)\in \D_2\\
1,& \text{if } \(r_1,r_2\)\in \D_3\\
0,& \text{if } \(r_1,r_2\)\in \D_4
\end{cases}\end{equation}
with $\bar{r}_i:=\alpha-r_j$ for $i=1,2$ and $j=3-i$. The value $\phi_2$ of Miner $2$ is analogous and the oceanic value $\Phi$ is simply equal to $\Phi=1-\phi_1-\phi_2$. \par
The interpretation of the values in the extreme regions $\D_3$ and $\D_4$ is straightforward. In $\D_3$, miner $1$ controls more than $50\%$ of the resources and hence, has absolute power over the blockchain. This implies that her value is equal to $1$ and consequently, the value for both miner $2$ and the oceanic miners is $0$. Region $\D_4$ is similar. The interesting cases arise whenever $\(r_1,r_2\)\in \D_2$, i.e., when the major miners and the ocean, each control less than $50\%$ of the resources, or $\(r_1,r_2\)\in \D_1$, i.e., when the resources controlled by the ocean account for more than half of the total resources. This case is also referred to as the \emph{interior case} in the original paper. Some instantiations in regions $\D_1$ and $\D_2$ are presented in \Cref{tab:interior}.

\begin{table}[!htb]
\vspace{-0.2cm}\centering
\begin{tabular}{ll*{3}{C{0.8cm}}c*{3}{C{0.8cm}}c*{3}{C{0.8cm}}}
\toprule
&& \multicolumn{3}{c}{Resources \%} && \multicolumn{3}{c}{Values \%} && \multicolumn{3}{c}{Ratios}\\
\cmidrule{3-5}\cmidrule{7-9}\cmidrule{11-13}
&& $r_1$ & $r_2$ & $\alpha$ &&  $\phi_1$ & $\phi_2$ & $\Phi$ && $v_1$ & $v_2$ & $\voc$\\
\midrule
$\D_1$: Interior Game 
&& 40 & 9 & 51  &&  65 & 4 & 31 && 1.62 & 0.42 & 0.62 \\
&& 30 & 19 & 51 && 37 & 15 & 48 && 1.23 & 0.81 & 0.94 \\
&& 25 & 24 & 51 && 26 & 24 & 50 && 1.04 & 1.00 & 0.98 \\
\midrule
$\D_2$: Balance of Power
&& 35 & 20 & 45 && 44 & 11 & 44 && 1.27 & 0.56 & 0.99 \\
&& 40 & 30 & 30 && 44 & 11 & 44 && 1.11 & 0.37 & 1.48 \\
&& 40 & 40 & 20 && 25 & 25 & 50 && 0.63 & 0.63 & 2.5 \\
\bottomrule
\end{tabular}\\[0.4cm]
\caption{Resources, values and values per unit of resource for various configurations in the $\D_1$ and $\D_2$ regions. The resources $r_i, i=1,2$ are selected arbitrarily, $\alpha=1-r_1-r_2$, the values $\phi, i=1,2$ and $\Phi$ are given by \eqref{ex:value} and the ratios $v_i, i=1,2$ and $\voc$ by \eqref{eq:ratio}.}
\label{tab:interior}\vspace{-0.6cm}
\end{table}
An indicative observation -- which does not aim to an exhaustive analysis of the above measurements -- is that the values and the ratios unveil disparities between \emph{shares} and actual \emph{influence} or \emph{power} of the participating entities. For example, there are instances, as in the $\(40,9,51\)$-configuration (first row in $\D_1$), in which a major miners' ratio is larger than the ratio of oceanic players. This imbalance generates a motive for oceanic players to merge with that miner to increase the power of their individual resources. Equivalently, the large miner has an increased influence to attract resources from the ocean. The picture is totally different in the $\(40,40,20\)$-configuration (third row of $\D_2$), in which the competition between the major miners raises the value of resources owned by the ocenic players. Both cases can be contrasted to the stability in the $\(25,24,51\)$-configuration, in which all $3$ ratios are approximately equal to $1$.\par
Yet, as argued in \cite{Mi78}, the interpretation of values should be done with caution and only in addition to complementary analytical tools. This is because values do not take into account qualitative factors such as ethical commitments, operational constraints or other kinds of incentives.
\newpage
\section{Individual Mining is not Stable}
\label{sec:results}
A direct outcome of applying the model of Oceanic Games in the blockchain context is the next result due to \cite{Mi78}. Both parts of \Cref{thm:main} make critical use of the assumption that the majority of mining resources is controlled by oceanic players. Their proof relies on a recursion in the number $m$ of major miners and can be found in \cite{Mi78}. Here, we will focus on the interpretation of \Cref{thm:main} and its application in blockchain context. 

\begin{theorem}[\cite{Mi78}] \label{thm:main} Let $\Gamma=[0.5;r_1,r_2,\dots,r_m;\alpha]$ with $M=\{1,2,\dots,m\}$ be a blockchain oceanic game, such that $r\(M\)< 0.5 \le \alpha$, i.e., such that the majority of mining resources is controlled by individual (oceanic) miners. Then
\begin{description}[leftmargin=*]
\item[(a)] The value $\phi_i$ of any major player $i\in M$ in $\Gamma$ is given by 
\[\phi_i=\frac{r_i}{\alpha^m}\sum_{S\subseteq M-\{i\}}\left[c_s\prod_{j\in S}r_j\prod_{k\notin S}\(\alpha-r_k\)\right]\]
where $c_s:=s!\left[\frac{1}{s!}-\frac{1}{\(s-1\)!}+\dots+\(-1\)^s\right]$ and $s:=|S|$ is the number of major miners in $S$. The oceanic value $\Phi$ is equal to $\Phi=1-\sum_{i\in M}\phi_i$.
\item[(b)] If $\Gamma^+=[0.5;r_1,r_2,\dots,r_m,r_{m+1};\alpha]$ for some $r_{m+1}>0$, and $\Phi^+,\phi^+_i, i=1,\dots,m+1$ are the values in $\Gamma^+$, then
\[\phi^+_{m+1}/r_{m+1}=\Phi/\alpha\] or equivalently, $v^+_{m+1}=\voc$. 
\end{description}
\end{theorem}

\begin{description}[leftmargin=0cm]
\item[Interpretation of \Cref{thm:main}.]
Statement (a) of \Cref{thm:main} is an analytical result which yields the exact formula to compute the values of the major miners and the ocean. Its usefuleness will become aparent in the applications. Statement (b) carries more intuition. It states that the value-per-unit of resource of a miner entering $\Gamma$ is equal to the oceanic value-per-unit of resource in $\Gamma$. One possible interpretation, also supported by \cite{Mi78}, is that this provides a stability argument in favor of decentralization, in the sense that there is no incentive for the formation of a \qt{cartel} or a mining pool, \emph{provided that} the size of the ocean is big enough, i.e., provided that the ocean controls the majority of the resources\footnote{This statement actually holds for any quota $q\in\(0,1\)$ and not only for $q=0.5$ as formulated here.}. \par
However, as we will see in the following applications, this picture is misleading and decentralization is actually not stable. In practice, the oceanic value per unit of resource in $\Gamma^+$ can go below the value per unit of resource of the crystallizing or newly entering entity. Hence, \emph{given that} a set of miners can coordinate their actions, then it may be beneficial for them to either crystallize out of the ocean or to acquire exogenous resources and form in both cases a single mining entity. 
\end{description}

\subsection{Applications of \Cref{thm:main}}
The above interpretations of \Cref{thm:main} are illustrated via the simulation of two representative scenarios. In both cases, we assess the stability of initial distributions of mining resources, in which the majority of resources is controlled by the ocean. This is achieved by comparing the oceanic value per unit of resource to the value per unit of resource of the same miners when acting as single entity.  

\begin{description}[leftmargin=*]
\item[I. Crystallization out of the ocean:] \label{ex:crystallization}
In the first scenario, we consider an instance of the blockchain oceanic game in which all resources are initially controlled by oceanic players. This is described by the game $\Gamma=[50\%;\alpha=100]$ and $M=\emptyset$. Then, we simulate a gradual formation of a single mining entity by the process of \emph{crystallization} out of the ocean. This is captured by a sequence of games (instances) $\Gamma^+=[50\%;r_1;\alpha]$ with $0<r_1<50$ and $\alpha=100-r_1$. For each instance, we calculate the value per unit of resource of the single entity that is forming out of the ocean and compare it with the value per unit of oceanic resource. The results are shown in \Cref{fig:crystallization}. 

\begin{figure}[!htb]\vspace{-0.3cm}
\centering
\includegraphics[trim={0.35cm 0.3cm 0.3cm 0cm}]{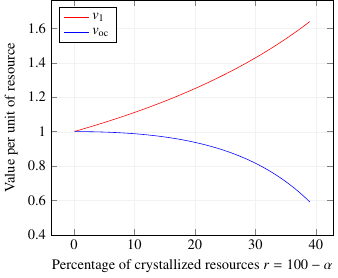}
\caption{The value per unit of resource of a single entity, $m=1$, that is forming by mergers (crystallization) out of the ocean (red line) and the value per unit of oceanic resource (blue line). The total percentage of resources that is controlled by the crystallizing entity is shown in the horizontal axis.}
\label{fig:crystallization}\vspace*{-0.4cm}
\end{figure}
It is apparent that $v_1$ is higher than $\voc$ even for arbitrarily low values of $r_1$ and that the difference is increasing in the percentage of crystallized resources. This uncovers a motive for coalition formations and merging between miners, even if the initial distribution is perfectly decentralized. Further simulations (not shown here) demonstrate that the same picture continues to hold even if $M\neq\emptyset$, as long as $\alpha>50\%$ and no single miner in $M$ holds a percentage close to $50\%$. If there exists a \qt{large} miner $i\in M$ with, e.g., $r_i>40\%$, then the oceanic players may be disincentivized to collude. However, this is only a semblance of stability, since in this case, oceanic miners have an incentive to merge with the \qt{large} miner.\\

%
%

\item[II. Acquisition of exogenous resources:]
In the second scenario, we consider miners who are acquiring exogenous resources to enter the mining process. We assume that these miners can either enter the ocean and mine individually or collude and form a single mining entity. We want to compare the value per unit of resource in these two cases. Formally, we denote the current distribution of resources by $\Gamma=[50\%; r_1,r_2,\dots,r_m;\alpha]$ with $\alpha>50\%$ and the total mining resources of the new entities by $w$. We want to compare 

\begin{itemize}[topsep=0.2cm]
\item $v_{m+1}^+:=\phi^+_{m+1}/w$ in the game $\Gamma^+{}:=[50\%; r_1,r_2,\dots,r_m, w;\alpha]$ to 
\item $\voc^{\text{o}}:=\Phi^\text{o}/\(\alpha+w\)$ in the game $\Gamma^{\text{o}}:=[50\%; r_1,r_2,\dots,r_m;\alpha+w]$.
\end{itemize}

The game $\Gamma^+$ describes the instance in which the entering miners merge in a single mining entity and the game $\Gamma^{\text{o}}$ the instance in which the entering miners become part of the ocean and mine individually. We assume that initially, the majority of resources is controlled by oceanic players and that there exist two other major mining entities. It turns out that the share of mining resources of the other major entities influences the incentives of the entering miners. To see this, we consider two cases.

\begin{description}[leftmargin=*,itemsep=5pt,topsep=0.2cm]
\item[Case 1:] Let $\Gamma=[50\%;6,4;90]$, so that $\Gamma^+=[50\%;6,4,w;90]$ and $\Gamma^\text{o}=[50\%; 6,$ $ 4; 90+w]$ for any $w>0$. As shown in \Cref{fig:addition}, in this case, both major miners are not large enough to create entry barriers for the third entity and $v^+_{m+1}>\voc^\text{o}$ for any $w>0$.
\begin{figure}[!htb]\vspace{-0.3cm}
\centering
\includegraphics[trim={0.35cm 0.3cm 0.3cm 0.1cm}]{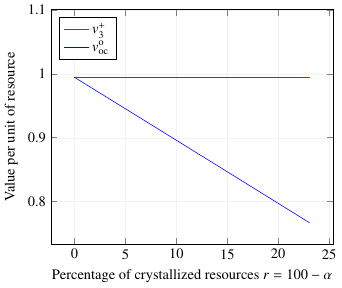}
\caption{The value per unit of resource of the entering miners when they enter as a single entity (red line) and their value per unit of resource when they enter as individual oceanic miners (blue line). The additional resources are shown as a percentage of the total resources in the horizontal axis.}
\label{fig:addition}\vspace*{-0.4cm}
\end{figure}
In agreement with \Cref{thm:main}(b), the value per unit of resource, $v_{3}^+$, of the new miners when they enter as a single entity is equal to the oceanic value $\voc$ in the initial game $\Gamma$ (red line). According to \cite{Mi78} this implies that \qt{there is no incentive for a new entity to form}. However, this only says half the truth. As we can see by the blue line, if the newly entering miners enter the ocean as individual miners, then their value per unit of resource will be lower compared to the case in which they collude. Hence, \emph{given that} a group of entering miners are capable to coordinate their actions, then they are better off if they enter as a single entity than as oceanic players. 
\item[Case 2:] Let $\Gamma=[50\%;55,5;90]$, so that $\Gamma^+=[50\%;55,5,w;90]$ and $\Gamma^\text{o}=[50\%; 55, 5; 90+w]$ for any $w>0$. As shown in \Cref{fig:addition2}, in this case, the presence of major miner $1$ seems to create a disincentive for a forming coalition and $v^+_{m+1}<\voc^\text{o}$ for any $w>0$ such that $w+90<50\%$.
\begin{figure}[!htb]\vspace{-0.3cm}
\centering
\includegraphics[trim={0.35cm 0.3cm 0.25cm 0.1cm}]{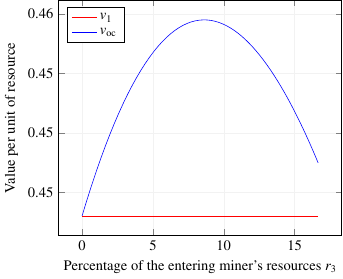}
\caption{The value per unit of resource of the entering miners when they enter as a single entity (red line) and their value per unit of resource when they enter as individual oceanic miners (blue line). The additional resources are shown as a percentage of the total resources in the horizontal axis.}
\label{fig:addition2}\vspace*{-0.4cm}
\end{figure}
The resulting picture shows that we cannot generalize the outcome of the previous case. In particular, we conclude that whether the entering miners have an incentive to form a single entity or to join the ocean as individual miners, may depend on the actual distribution of resources among the existing major miners and the ocean. However, this is only a semblance of stability, stemming from an already centralized initial distibution ($r_1>33\%$). In this case, oceanic miners actually have a stronger incentive to merge with miner $1$ instead of forming a new entity. 
\end{description}
\end{description}
The previous simulations create an inconclusive picture. In general, the incentives for miners to merge seem to depend on the current distribution of resources. Since blockchain mining is a dynamical system that evolves over time, they suggest that even if the blockchain starts from a sufficiently decentralized point, then it is unlikely to remain decentralized also in the future or equivalently that decentralization creates a \emph{negative feedback loop}, \cite{Ze00,Ha02}. The dynamics of the coalition formation process and the entry barriers resemble these of conventional economic markets of either perfect or oligopolistic competition. These findings provide an alternative perspective to cryptocurrency mining along with \cite{Ar19}, and suggest the need for further research in this direction. 

\begin{description}[leftmargin=0cm]
\item[III. Revisiting Bitcoin and Ethereum:]\par
Finally, we return to the current distribution of resources in Bitcoin and Ethereum and estimate the values per unit of resource for each mining entity,  cf. \Cref{tab:hashpower}. For this, we use the publicly available \href{http://homepages.warwick.ac.uk/~ecaae/ssocean.html}{ssocean} software. The results are presented in \Cref{fig:current}.

\begin{figure}[!htb]
\centering
\includegraphics[trim={0.35cm 0.3cm 0.25cm 0.1cm}]{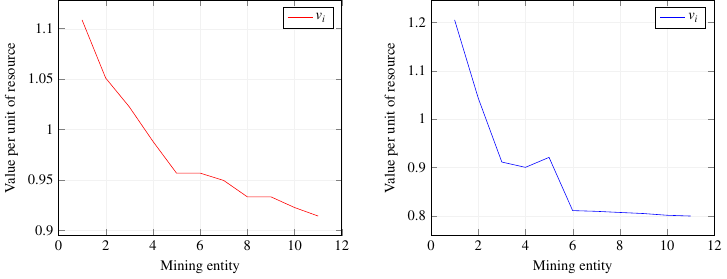}
\caption{Value per unit of resource of current Bitcoin (red, left panel) and Ethereum (blue, right panel) mining entities: big miners and the ocean.}
\label{fig:current}
\end{figure}

The horizontal axis represents the mining pools and the ocean as given in \Cref{tab:hashpower} in descending order, i.e., with \qt{1} corresponding to the largest miner and \qt{11} to oceanic miners. The results are revealing: excluding a minor knick in the Ethereum graph, the largest the mining pool, the largest the value per unit of resource. 
\end{description}

\section{General Issues, Research Perspectives \& Limitations}
\label{sub:research}

The application of the oceanic-game model in blockchain mining opens new research perspectives but also has its own limitations. Beyond the insight from existing results, a complete model needs to account for the additional challenges and address the questions that are specific to the blockchain context. In the following discussion, we raise such relevant issues, discuss their connection and research possibilities via the current model and identify potential limitations.
\begin{description}[itemsep=0.15cm, leftmargin=0cm]
\item[Cryptocurrencies as Resources:] 
The difference between PoW and PoS in terms of their resources -- computational power versus native coins -- has a direct impact on both the mining process and the value of the underlying cryptocurrency \cite{Kok20}. When coins are used as mining resources (PoS protocols), their value depends on their distribution among existing miners, their availability for prospective miners and the returns (profits) from mining. This in contrast to PoW protocols, in which the price of the resources -- e.g., hardware and electricity -- is not tied to the price of the underlying cryptocurrency.
\item[Resource Acquisition \& Entry Barriers:]
The above suggest that the nature of protocol resources may also generate different entry barriers. In PoS, the acquisition of protocol resources, i.e., coins, by prospective nodes depends on the willingness of current owners to exchange their coins and the way that new coins are minted. Different configurations may lead to high entry barriers and centralization. In PoW, computational power is essentially unlimited and acquisition of additional resources is independent of the underlying cryptocurrency. This implies lower entry barriers but also more frequent changes in the configuration (distribution) of resources among miners.\par
In particular, the current cost of acquiring enough computational power to control the majority in the Bitcoin and Ethereum PoW-blockchains is estimated at 1.5 billion US Dollars \cite{Bo18}. This amount is well within the budget of several physical or legal entities worldwide. Moreover, it is \emph{independent} of the value of the underlying cryptocurrency and depends only on the size of the network and the hardware and electricity costs. With this in mind, it is natural to ask: how stable are PoW blockchains against arbitrary authorities able to acquire the majority of resources? How relevant are these questions to the current distribution of resources and how do they translate in the PoS setting? \par
In this context, further work on blockchain oceanic games can aid the community to raise and study questions about investment in cryptocurrencies. When viewed not only as assets but also as means to gain power in the mining process and the governance of a blockchain, cryptocurrencies fit to the current perspective and their mechanics can be better understood.
\item[Mathematical Modelling:]
From a mathematical perspective, oceanic games bridge the gap between atomic and non-atomic congestion games \cite{Kl09}. Yet, the use of \emph{values} instead of equlibria to study real settings has its own limitations \cite{Mi78,Sh88}. This is mainly due to the probabilistic derivation of values, which ignores qualitative aspects such as ethical commitments, preferences or any other motives of the participating agents. However, despite these limitations, if properly interpreted, values can become a powerful tool in the analysis of strategic interactions. In an immediate direction, they can be used to rethink the notion of \emph{blockchain fairness} or \emph{equitability}, which is currently based on the theoretically tentative premise that \emph{one unit of resource -- one vote} also implies fairness \cite{Fa18,Wo89}.
\end{description}

\section{Conclusions}
\label{sec:conclusions}
In this paper, we employed the concept of Oceanic Games \cite{Mi78}, to model and study strategic interactions in blockchain mining. Oceanic Games have been used in conventional economics to analyze decision making in corporate settings with a small number of major players  -- shareholders or here, mining pools -- and a continuum of minor, individually insignificant players, called the \emph{ocean}. This stream of literature focuses on the measurement of the value per miner and per unit of resource for each miner given a distribution of resources. Values are then interpreted as strategic components in decisions related to resource acquisition, mergers and coalition formations and offer an alternative perspective to the common equilibration models.\par
An immediate implication of existing results was that given a sufficiently large initial distribution of resources, there are \emph{incentives} both for active and for newly entering miners to merge (form cartels or coalitions) and act as single entities. These observations provide an alternative justification of the observed centralization and concentration of power in the mining process of the major cryptocurrencies. Contrary to common perceptions, they amount to the existence of a negative feedback loop in terms of decentralization as a core ingredient in permissionless blockchain philosophy, \cite{Leo18}, and reveal the need for futher research in this direction. In a general discussion, we identified critical issues related to resource acquisition, entry barriers and centralization risks in blockchain mining and formulated relevant questions that may be answered by further exploration of the present model. These findings can be placed in the broader context of governance and long-term sustainability for pemissionless blockchains.

\section*{Acknowledgements}
Stefanos Leonardos thanks his PhD advisor, Costis Melolidakis, for introducing him to Oceanic Games. Stefanos Leonardos and Georgios Piliouras acknowledge that this work was supported in part by the National Research Foundation (NRF), Prime Minister's Office, Singapore, under its National Cybersecurity R\&D Programme (Award No. NRF2016NCR-NCR002-028) and administered by the National Cybersecurity R\&D Directorate. Georgios Piliouras acknowledges SUTD grant SRG ESD 2015 097, MOE AcRF Tier 2 Grant 2016-T2-1-170 and NRF 2018 Fellowship NRF-NRFF2018-07. Finally, the authors would like to thank Dennis Leech for making publicly the \href{}{ssocean} software that was used in application III of the present paper.

\bibliographystyle{splncs04}
\bibliography{oceanic_games_bib}
\end{document}